\documentclass[pre,twocolumn,fleqn,showpacs,showkeys,amsmath]{revtex4}
\usepackage{graphicx}
\usepackage{amssymb}

\newcommand{\rme}{\mathrm{e}}

\newcommand{\rmd}{\mathrm{d}}

\begin{document}

\setcounter{page}{1}%

\title{ASYMMETRIC RANDOM WALK IN A ONE-DIMENSIONAL MULTI-ZONE ENVIRONMENT}

\author{A.V.~Nazarenko}\email{nazarenko@bitp.kiev.ua}
\affiliation{Bogolyubov Institute for Theoretical Physics of NAS of Ukraine, \\ 
4b, Metrologichna Str., Kyiv 03680, Ukraine}%

\author{V.~Blavatska}\email{viktoria@icmp.lviv.ua}
\affiliation{Institute for Condensed Matter Physics of NAS of Ukraine,\\ 
1, Svientsitskii Str., Lviv 79011, Ukraine}


\begin{abstract}
We consider a random walk model in a one-dimensional environment, formed by several zones of finite width
with the fixed transition probabilities. It is also assumed that the transitions to the left
and right neighboring points have unequal probabilities. In continuous limit, we derive analytically the
probability distribution function, which is mainly determined by a walker diffusion and drift and accounts
perturbatively for interface effects between zones. It is used for computing the probability to find a
walker in a given space-time point and the time dependence of the mean squared displacement of a walker,
which reveals the transient anomalous diffusion. To justify our approach, the probability function is
compared with the results of numerical simulations for a three-zone environment.
\end{abstract}



\maketitle

\section{Introduction}

The present paper is devoted to a deeper study of a random walk (RW) in a one-dimensional inhomogeneous
environment, consisting of $N$ zones with constant parameters. The base model is formulated in our previous
work~\cite{NaBla}, where we have referred to a number of problems~\cite{Ascher60,Lehner,Gupta,Tanner,Percus,BHMST,NFJH,Powels}
investigated before and influencing our motivation. We have tried there to apply an analytical
approach for finding the probability distribution function (PDF) of a walker in heterogeneous environment.

Here, we supply the RW model by seating an attractor/repulsor at the root point, which serves as
a source of external field. Its presence determines the preferable directions of a walk and leads therefore
to emergence of inequality (asymmetry) between probabilities of transition to the left and to the right.

Our investigations are stimulated by the RW models application in the majority of fields like polymer physics,
economics, computer sciences~\cite{RWbook}. Besides it, RW is often used as simple mathematical formulation
of diffusion process. Moreover, the RW in inhomogeneous environment is of great interest because of its
connection with transport phenomena in fractures and porous rocks, diffusion of particles in gels, colloidal
solutions and biological cells (see, e.g., \cite{Havlin87} for a review). 

Thus, we formulate here the RW in $N$-zone environment, located along coordinate axis and symmetric
under coordinate sign inversion. The transition probability is assumed to be varying within the different zones
and unequal for the left and right steps.

This model is engaged to describe, for instance, the linear chaotic structures, passing through the finite number
of zones with viscous properties in external field. Imagining for a moment a few-dimensional problem, these may be
polymer chains crossing the cellular membranes \cite{MGCh}. Applying an analytical approach, we take a possibility
here to evaluate the mean squared end-to-end distance in a one dimension.

We can also suppose the model applicability to a study of chaotic particle dynamics with barriers, which are
associated with various matter sorts or (static) properties of space itself. Our goal is to derive also
the probability as a function of space-time, which is accounting for transition between zones
due to diffusion and advection.

In principal, we have no preference for the treatments and investigate the general properties. We are interested
in finding new effects and functions in comparison with \cite{NaBla}, approaching our model step by step to realistic
situations which can be considered elsewhere.

Analytically, we are aiming to reduce the problem in continuous limit to finding a PDF from differential equation
with a diffusion coefficient and a drift velocity, which inherit the established step-like space dependence.
Further, a space-time evolution might be characterized by the averages, which are found by means of PDF and allow
us to compare the properties of similar models with various parameters. We would like also to demonstrate
a strong dependence of the meaningful quantities on attractive or repulsive character of the starting point.
In particular, it is predicted a steady state existence when the diffusion and the advection are equilibrated.

As it is already shown in \cite{NaBla}, there is a possibility to compare the derived probability function
to find a walker in a given point of space-time with the result extracted from numerical simulations on the lattice.
We test here these functions in the case of three-zone environment too.

Technically, a time dependence of the mean squared displacement does not look easily computable. Then, we neglect
the interface effects, but account accurately for the bulk ones. This simplification is justified by studies,
performed without advection (drift) in \cite{NaBla}. There, the time asymptotic of the variance corresponds to
uniform model, while the multi-zone structure of environment leads to emergence of transient anomalous diffusion
at finite time. These phenomena are investigated here in the drift presence.

The layout of the paper is as follows. In the next Section, we fix the random walk rules and obtain
the differential equation for PDF. An approximate PDF is found in Section~3. The probability
function is calculated analytically in Section~4 and is compared with the numerical simulations performed.
After computing the mean squared displacement in Section~5, we end up with giving discussion and outlook.

\section{Asymmetric RW in $N$-Zone Environment}

We start with a lattice RW model, considered as a Markov process with either zeroth or unitary step
in space after successive unitary step in time. The walk is determined here by stationary
transition probability $T(x_{t+1},x_t)$ defined as
\begin{equation}
T(x,y)=p(y)\delta_{x-y,-1}+q(y)\delta_{x-y,1}+r(y)\delta_{x-y,0},
\end{equation}
where $\sum_{x\in{\mathbb Z}}T(x,y)=p(y)+q(y)+r(y)=1$ is implied for a fixed $y\in{\mathbb Z}$;
$\delta_{x,y}$ is the Kronecker symbol.

Functions $p(x)$ and $q(x)$ determine probabilities to find a walker at points $x-1$ and $x+1$ if it
was at $x$ in a previous time instant, respectively; $r(x)$ corresponds to probability of adhesion
(adsorption) \cite{Gupta}. In general, $p(x)$, $q(x)$ are regarded as arbitrary non-negative functions
less than the half of unit.

Let sequence $\{a_n\}$ of $N$ positive numbers,
\begin{equation}
0=a_0<a_1<\ldots<a_{N-1}<a_N=\infty,
\end{equation}
define the separation points of environment zones in space $x\in\mathbb{Z}_+$.

Reproducing the same configuration for negative $x$ by inverting $a_n\to-a_n$, we introduce
the characteristic functions $\chi_n$ of zones $[-a_n;-a_{n-1}]\cup[a_{n-1};a_n]$: $\chi_n(x)=1$
for $|x|\in(a_{n-1};a_n)$, $\chi_n(\pm a_n)=\chi_n(\pm a_{n-1})=1/2$ to obtain always the arithmetic
mean of the left and right-handed functions at separation points, and $\chi_n(x)=0$ otherwise.
To make use these in differential calculus, $\chi_n(x)$ are written as the distributions:
\begin{eqnarray}
\chi_1(x)&=&\theta(a_1-|x|),\nonumber\\
\chi_n(x)&=&\theta(a_n-|x|)-\theta(a_{n-1}-|x|),\quad n>1,
\end{eqnarray}
Here $\theta(x)=[1+\mathrm{sign}(x)]/2$ is the Heaviside function.

The functions $\{\chi_n\}$ are orthogonal:
\begin{eqnarray}
&&\sum\limits_{n=1}^N\chi_n(x)=1,\quad |x|<a_N;\\
&&\chi_n(x)\chi_m(x)=0\quad {\rm for}\quad n\not=m,\quad x\not=\{\pm a_n\}.\label{cond2}
\end{eqnarray}

Given the basis, we further introduce $N+1$ constant parameters $\{d_n;V\}$, where $d_n\leq1/2$ defines
the diffusion coefficient of $n$-th zone; $V$ is a measure of probability asymmetry
and plays a role of external field. Supposing the drift velocity dependence on both attractor strength
and zone properties, we define $n$-th zone velocity as $v_n=V\sqrt{d_n}$.

Thus, we consider the asymmetric RW in heterogeneous environment with the following rules:
\begin{eqnarray}
&&p(x)=\sum\limits_{n=1}^N\left(d_n-\frac{1}{2}V\sqrt{d_n}\varepsilon_x\right)\chi_n(x),\nonumber\\
&&q(x)=\sum\limits_{n=1}^N\left(d_n+\frac{1}{2}V\sqrt{d_n}\varepsilon_x\right)\chi_n(x),\nonumber\\
&&r(x)=1-2\sum\limits_{n=1}^N d_n\chi_n(x),
\label{rules}
\end{eqnarray}
where $\varepsilon_x\equiv\mathrm{sign}(x)$.

To preserve a probability meaning of functions $p(x)$ and $q(x)$, we require $0<2d_n\pm V\sqrt{d_n}\leq1$.

The rules (\ref{rules}) allow us to simulate immediately the random trajectories for various sets $\{a_n;d_n;V\}$,
which we analyze here analytically.

Physically, definition of $v_n$ leads to a common and finite time $\tau=1/V^2$ of advection residence
in a whole space. Therefore, the $n$-th zone advection length $L^\mathrm{a}_n\simeq|v_n|t$ and
the diffusion length $L^\mathrm{d}_n\simeq\sqrt{d_nt}$ give us the Peclet number~\cite{Pat}
$\mathrm{Pe}\equiv(L^\mathrm{a}_n/L^\mathrm{d}_n)^2=t/\tau$ for all $n$. Thus, the both processes are
tantamount at $t$ obeying $\mathrm{Pe}\sim 1\div10$ as usual. At larger $t$, the advection becomes
dominant.

Although we give a common analytical description at $V>0$ and $V<0$, these cases physically differ
because of repulsive or attractive role of the origin, where velocity sign is inverted due to
$\varepsilon_x$. Scenario with $V<0$ accords with the presence of a single attractor at $x=0$,
while the case of $V>0$ admits an existence of two attractors at $x=\pm\infty$,  which produce
two particle flows, moving in opposite directions.

The Markovian evolution of the probability distribution function (PDF) $P(x,t)$, so that
$P(x,0)=\delta_{x,0}$, can be given by master equation with arbitrary distance $\ell$ and time $\tau$
between successive steps as
\begin{eqnarray}
P(x,t+\tau)&=&r(x)P(x,t)+p(x+\ell)P(x+\ell,t)\nonumber\\
&&+q(x-\ell)P(x-\ell,t).
\label{ME}
\end{eqnarray}

To obtain a differential equation at large $t$, (\ref{ME}) is expanded into the Taylor series up
to the order $O(\ell^2,\tau)$. Omitting the rest terms, the lattice parameters are fixed to give
constant scales $\ell/\tau=1$ and $\ell^2/2\tau=1/2$ of the drift velocity and diffusivity, respectively.

Further, defining the functions for all $d_n>0$,
\begin{equation}
D(\alpha,x)=\sum\limits_{n=1}^N (d_n)^{\alpha}\chi_n(x),\quad
D(x)\equiv D(1,x),
\end{equation}
we arrive in continuous limit at differential equation of RW in diffusion approximation:
\begin{eqnarray}
\partial_tP(x,t)&=&\partial^2_x[D(x)P(x,t)]
\nonumber\\
&&-V\partial_x\left[\varepsilon_xD(1/2,x)P(x,t)\right],
\nonumber\\
P(x,0)&=&\delta(x).
\label{idiff}
\end{eqnarray}

Differentiating, we can use $D^{\prime}(\alpha+\beta,x)=D^{\prime}(\alpha,x) D(\beta,x)
+D(\alpha,x) D^{\prime}(\beta,x),$ where the prime means derivative with respect
to coordinate \cite{NaBla}.

Note also that $D(\alpha,x)D(\beta,x)=D(\alpha+\beta,x)$ for $x\in\mathbb{R}\backslash\{\pm a_n\}$,
and $D(\alpha,\pm a_n)\not=[D(\pm a_n)]^{\alpha}$ in general.

Evolving in space-time, the PDF defines the normalized statistical measure $\mu_t$ for a fixed $t$:
\begin{equation}
\rmd\mu_t=P(x,t)\rmd x,\qquad
\int\rmd\mu_t=1,
\end{equation}
which is used for computing the averages.

\section{Finding a PDF}

To find a PDF, we follow~\cite{NaBla} and concentrate the geometrical data in the new coordinate
\begin{equation}\label{yfun}
\xi(x)=\int_0^{x}D(-1/2,x^{\prime})\rmd x^{\prime}.
\end{equation}
Note that the derivatives of $\xi(x)$ are singular, in general, at the points $x=\{\pm a_n\}$.

Integrating (\ref{yfun}), we obtain $\xi(x)=\varepsilon_x X(-1/2,x)$,
\begin{eqnarray}
&&X(\alpha,x)\equiv\frac{1}{2}\sum\limits_{n=1}^N(d_n)^{\alpha}\left[l_n(x)-l_{n-1}(x)\right],
\label{Yfunc}\\
&&l_n(x)=a_n-||x|-a_n|.
\nonumber
\end{eqnarray}

We also define the functions ${\tilde D}(\alpha,\xi(x))=D(\alpha,x)$:
\begin{equation}
{\tilde D}(\alpha,\xi)=\sum\limits_{n=1}^N(d_n)^{\alpha}\tilde\chi_n(\xi),
\end{equation}
here $\tilde\chi_n(\xi)=\theta(b_n-|\xi|)-\theta(b_{n-1}-|\xi|)$ and $b_n\equiv \xi(a_n)$.

Then, introducing the probability distribution ${\cal P}(\xi,t)$, we re-write statistical measure  as
\begin{eqnarray}
\rmd\mu_t&=&{\cal P}(\xi,t)\rmd\xi
\nonumber\\
&=&{\cal P}(\xi(x),t) D(-1/2,x)\rmd x.
\end{eqnarray}

Substituting the re-defined $P(x,t)$ into (\ref{idiff}), we arrive at equation:
\begin{equation}\label{DP}
\partial_t{\cal P}+V\partial_\xi(\varepsilon_\xi{\cal P})
-\partial^2_\xi{\cal P}=\kappa\partial_\xi(\beta{\cal P}),
\end{equation}
where constant $\kappa$ controls an interface effect between zones; ${\cal P}(\xi,0)=\delta(\xi)$.

Reformulating the model in the terms of $\xi$, $V$ is regarded as
a global velocity, which takes the opposite signs in two infinite intervals of $\xi$: $\xi>0$
and $\xi<0$.

The right hand side of (\ref{DP}) can be reduced to the form with 
$\beta(\xi)={\tilde D}(-1/2,\xi)\partial_\xi{\tilde D}(1/2,\xi)$ and $\kappa=1$. Computations performed
lead to the expression:
\begin{equation}
\beta(\xi)=\varepsilon_\xi\sum\limits_{n=1}^{N-1}\beta_n\delta(|\xi|-b_n),
\ \ \beta_n=\frac{d_{n+1}-d_n}{2\sqrt{d_nd_{n+1}}}.
\label{beta}
\end{equation}
A sign of $\beta_n$ is defined by difference $d_{n+1}-d_n$, although the form of $\beta_n$ can vary.

We substitute now a formal series in $\kappa$:
\begin{equation}\label{sersol}
{\cal P}(\xi,t)=\varphi(\xi,t)+\sum\limits_{r=1}^{\infty}\kappa^r{\cal S}_r(\xi,t),
\end{equation}
where
\begin{eqnarray}\label{phi}
\varphi(\xi,t)&=&\frac{1}{\sqrt{4\pi t}}\exp{\left(-\frac{(|\xi|-Vt)^2}{4t}\right)}
\nonumber\\
&&-\frac{V}{4}\rme^{V|\xi|}\mathrm{erfc}\left(\frac{|\xi|+Vt}{2\sqrt{t}}\right)
\end{eqnarray}
is a basic and normalized solution to (\ref{DP}) at $\kappa=0$.

It is instructive to compare the fundamental solution $\theta(t)\varphi(\xi,t)$ with
analytical solutions to the advection-diffusion equation under other conditions from
recent works~\cite{GPSG,KJK,Singh}. 

Remaining problem is to examine the surface effect induced by $\beta(\xi)$, which disappears
in the homogeneous environment.

{In general, each term ${\cal S}_r$ is determined by divergence $\partial_\xi(\beta{\cal S}_{r-1})$
with a point-like carrier in the right-hand side of (\ref{DP}) and, therefore, results in
\begin{equation}
\int_0^\infty{\cal S}_r(\xi,t)\rmd\xi=0,\qquad r\geq1,
\end{equation}
what preserves the normalization. This integral statistically means that ${\cal S}_r$ is fluctuating
and a sign alternating function of space. Its magnitude is not suppressed by factor $\kappa^r$,
and the probability may fall down to negative value. Although the series (\ref{sersol}) allows us to
calculate all of ${\cal S}_r$ in a simple way, appropriate resummation is still needed.}

Nevertheless, we compute here the first-order term ${\cal S}_1$ from inhomogeneous equation
\begin{equation}\label{eqS}
\partial_t{\cal S}_1
+V\partial_\xi(\varepsilon_\xi{\cal S}_1)-\partial^2_\xi{\cal S}_1=\partial_\xi(\beta\varphi),
\end{equation}
contracting $\varphi(\xi,t)$ with $\partial_\xi(\beta\varphi)$ to obtain
\begin{eqnarray}
&&{\cal S}_1(\xi,t)=\sum\limits_{n=1}^{N-1}\sum\limits_{\epsilon=\pm}\epsilon\beta_n
\mathrm{sign}(\xi-\epsilon b_n)I^\epsilon_n(\xi,t),\\
&&I^\pm_n(\xi,t)=\int_0^tf(\xi\mp b_n,t-\tau)\varphi(b_n,\tau)\rmd\tau,
\end{eqnarray}
where $f(\xi,t)=\partial_{|\xi|}\varphi(\xi,t)$; $b_n>0$.

Performing an integration, one has
\begin{eqnarray}
&&\hspace*{-4mm}
I^\pm_n(\xi,t)=-\frac{1+V^2t}{4\sqrt{\pi t}}\exp{\left(-\frac{(|\xi\mp b_n|+b_n-Vt)^2}{4t}\right)}
\nonumber\\
&&\hspace*{2mm}
+\frac{V}{8}\rme^{V(|\xi\mp b_n|+b_n)}\mathrm{erfc}\left(\frac{|\xi\mp b_n|+b_n+Vt}{2\sqrt{t}}\right)
\nonumber\\
&&\hspace*{2mm}
\times[3+V(|\xi\mp b_n|+b_n)+V^2t].
\end{eqnarray}

Limiting ourselves by accounting for the first-order correction, we obtain our main result for PDF:
\begin{eqnarray}
P(x,t)&=&\varphi(\xi(x),t)D(-1/2,x)
\nonumber\\
&&+\kappa\sum\limits_{n=1}^{N-1}\beta_n[\mathrm{sign}(x-a_n)I^+_n(\xi(x),t)
\nonumber\\
&&-\mathrm{sign}(x+a_n)I^-_n(\xi(x),t)]D(-1/2,x),
\label{dist}
\end{eqnarray}
where $\mathrm{sign}(x\pm a_n)=\mathrm{sign}(\xi(x)\pm b_n)$.

At the vanishing $V$, this reproduces the result of \cite{NaBla} and describes the ordinary RW at
$2d_n=1$ for all $n$, when $\xi(x)=\sqrt{2}x$ and $\beta_n=0$.

If $d_{n+1}=d_n$ for two neighboring zones, the interface points $x=\pm a_n$ become regular,
and (\ref{dist}) welds these zones automatically.

Note that the correction by $\kappa$ looks improper at $V<0$ because it has no static limit at
large $t$. Although the diffusion in an attractor presence at $x=0$ has to be a relaxation process
leading to a steady state distribution $P_\mathrm{eq}(x)$ and preventing a collapse
$\langle x^2\rangle=0$ due to chaotic motion. Taking $t\to\infty$ and $\kappa=0$ ($\beta_n=0$)
in (\ref{dist}), one obtains
\begin{equation}
P_\mathrm{eq}(x)=\frac{|V|}{2}\exp{(-|V||\xi(x)|)}D(-1/2,x).
\end{equation}
It results in the limiting value of the variance $\langle x^2\rangle$ at
$t\to\infty$ as we shall see. Thus, the PDF (\ref{dist}) is applicable to the models with $V<0$
at $\kappa=0$.

Considering the case of $V>0$, the surface term is relevant at $|\beta_n|\ll1$ because $I^\pm_n$
is of the same order of magnitude as $\varphi$. As the result, Fig.~1 demonstrates two peaks of PDF,
tending to escape in opposite directions to infinity with increasing $t$. However, the speeds of
peak motion for two sets of parameters look different because of a different damping, caused by
adhesion.

\begin{figure}
\begin{center}
\includegraphics[width=8cm]{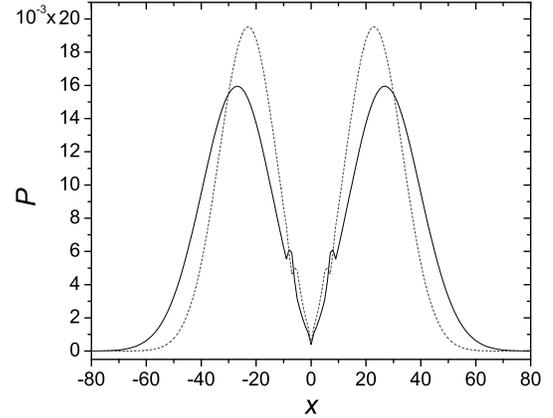}
\end{center}
\caption{Probability distribution function at $t=200$ and $V=0.2$ for two sets of environment parameters.
Solid curve corresponds to the model with $2d_1=0.6$, $2d_2=0.4$, $2d_3=0.9$; the dashed one is
for $2d_1=0.4$, $2d_2=0.9$, $2d_3=0.6$. In the both cases, $a_1=6$, $a_2=8$.}
\end{figure}

Note finally that Eq.~(\ref{DP}) can be transformed into the potential form by excluding
the term $\partial_\xi{\cal P}$.

\section{The Probability Function}

Using the PDF, let us compute the probability of walker manifestation at a given space-time point,
\begin{equation}\label{defPr}
{\rm Pr}(x,t)=\frac{1}{t}\int_0^tP(x,\tau)\rmd\tau,
\end{equation}
that is the frequency of visiting a point $x$ in a time $t$.

Integrating, $\varphi(\xi,t)$ results in
\begin{eqnarray}\label{Phi}
&&\Phi(\xi,t)=\frac{1}{\sqrt{4\pi t}}\exp{\left(-\frac{(|\xi|-Vt)^2}{4t}\right)}
\nonumber\\
&&+\frac{1}{4Vt}\left[\mathrm{erfc}\left(\frac{|\xi|-Vt}{2\sqrt{t}}\right)
-\rme^{V|\xi|}\mathrm{erfc}\left(\frac{|\xi|+Vt}{2\sqrt{t}}\right)\right]
\nonumber\\
&&-\frac{|\xi|+Vt}{4t}\rme^{V|\xi|}\mathrm{erfc}\left(\frac{|\xi|+Vt}{2\sqrt{t}}\right).
\end{eqnarray}

Computing the total probability, we arrive at
\begin{eqnarray}
\mathrm{Pr}(x,t)&=&\Phi(\xi(x),t)D(-1/2,x)+
\nonumber\\
&+&\kappa\sum\limits_{n=1}^{N-1}\beta_n[\mathrm{sign}(x-a_n){\cal I}^+_n(\xi(x),t)
\nonumber\\
&-&\mathrm{sign}(x+a_n){\cal I}^-_n(\xi(x),t)]D(-1/2,x),
\label{Prr}
\end{eqnarray}
where the first-order term in $\kappa$ is determined by
\begin{eqnarray}
&&\hspace*{-4mm}
{\cal I}^\pm_n(\xi,t)={\cal I}(|\xi\mp b_n|+b_n,t),\\
&&\hspace*{-4mm}
{\cal I}(\xi,t)=-\frac{1}{\sqrt{4\pi t}}\rme^{-(\xi-Vt)^2/(4t)}
\left(1+\frac{V\xi+V^2t}{4}\right)
\nonumber\\
&&\hspace*{-1mm}
+\frac{V}{4}\rme^{V\xi}\mathrm{erfc}\left(\frac{\xi+Vt}{2\sqrt{t}}\right)
\left[\frac{3}{2}+\frac{\xi}{Vt}+\frac{(\xi+Vt)^2}{4t}\right].
\end{eqnarray}

\begin{figure}[htbp]
\begin{picture}(80,45)
\put(-60.5,-110){\includegraphics[width=7.5cm,angle=0]{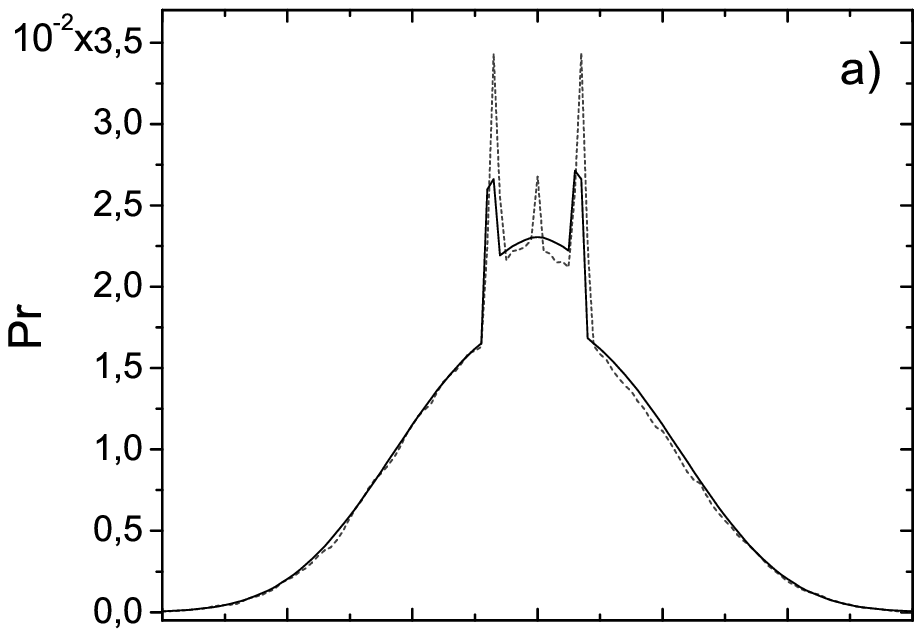}}
\put(-62.2,-260){\includegraphics[width=7.76cm,angle=0]{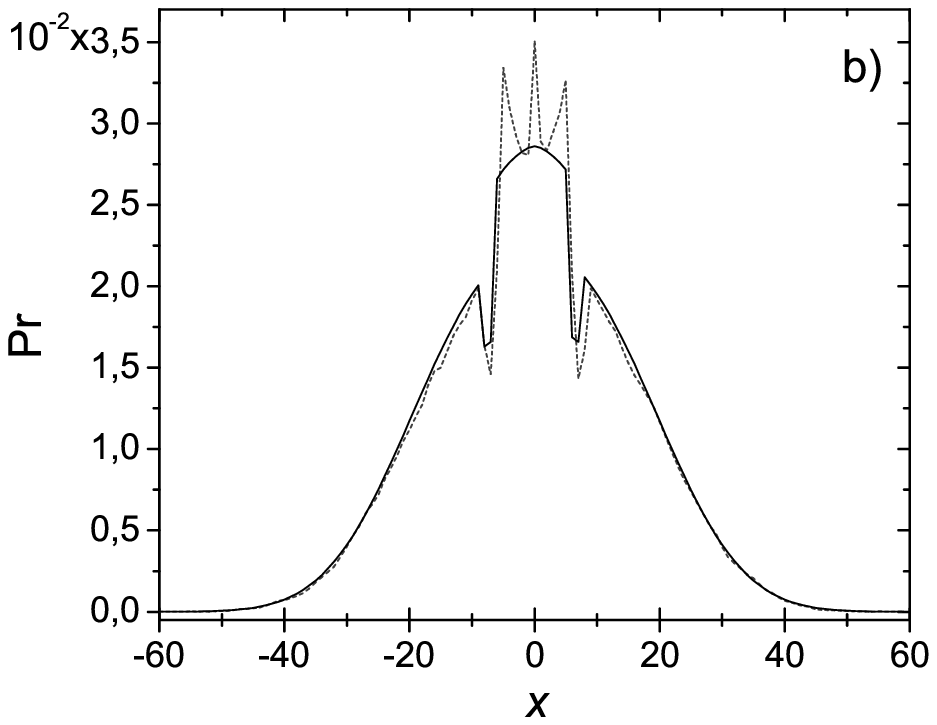}}
\end{picture}
\vspace*{90mm}
\caption{\label{fig2} Probability function at $t=200$ and $V=0.2$ for three-zone environment models.
Solid lines are described by (\ref{Prr}). Dashed ones result of averaging over 3000
random trajectories. Panel a) $2d_1=0.6$, $2d_2=0.4$, $2d_3=0.9$; b) $2d_1=0.4$, $2d_2=0.9$, $2d_3=0.6$.
In the both cases, $a_1=6$, $a_2=8$.}
\end{figure}

We test our formulas by comparing $\mathrm{Pr}(x,t)$ with outcomes of numerical simulations. Data are presented
in Fig.~2 and Fig.~3, putting respectively $\kappa=1$ and $\kappa=0$, as it was argued above. Indeed,
the bulk analytical solution is justified by RW simulation in three-zone environment. However, we hope that
the accounting for the series in $\kappa$ or its finite part, at least, will permit us to reproduce better
the considerable changes of the probability profile at short $\Delta x$.

The difference between peak heights obtained analytically and numerically at $x=0$ is independent on
approximation in $\kappa$ and can be also explained by the feature of our formalism, describing mainly
a walker behavior in the bulk .

At $V<0$, we see actually coincidence of three probability functions inside zones: numerical one,
$\mathrm{Pr}(x,t)$ at $t=200$, and $\mathrm{Pr}_\mathrm{eq}(x)=P_\mathrm{eq}(x)$ at $t\to\infty$.

\begin{figure}[htbp]
\begin{picture}(80,45)
\put(-60.5,-113){\includegraphics[width=7.5cm,angle=0]{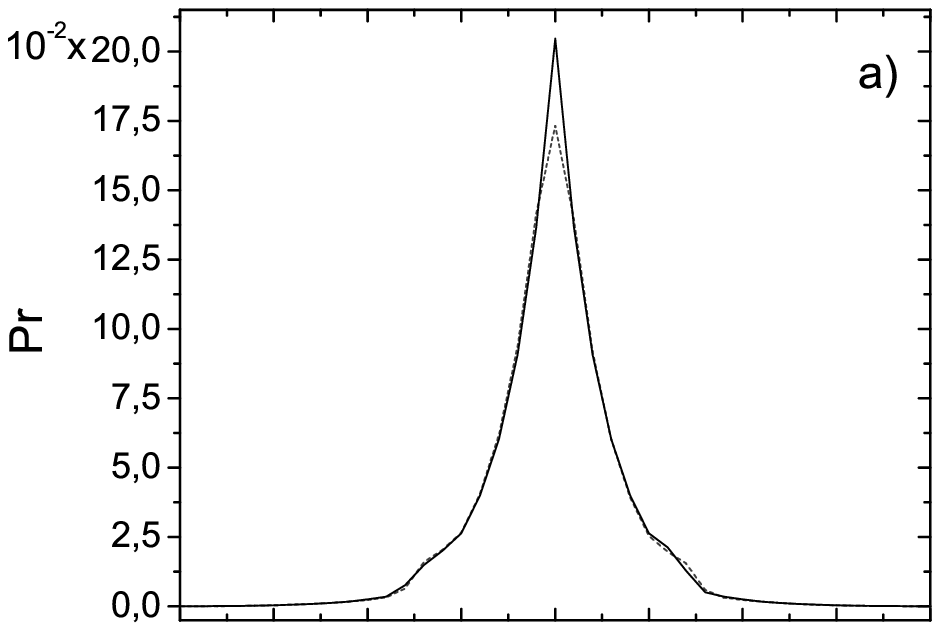}}
\put(-57,-260){\includegraphics[width=7.57cm,angle=0]{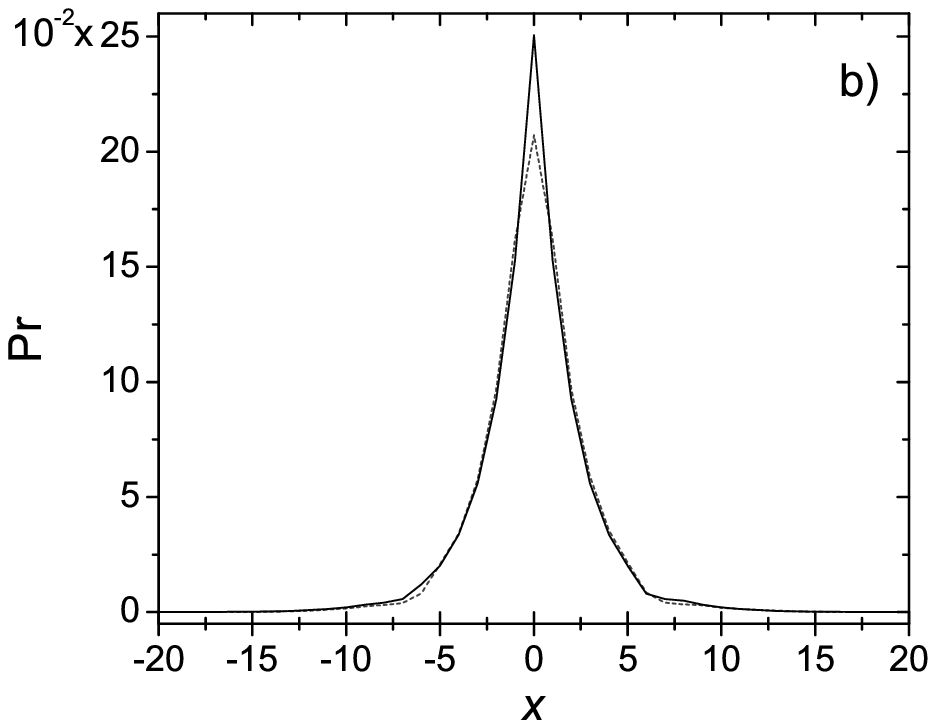}}
\end{picture}
\vspace*{90mm}
\caption{\label{fig3} Probability function at $t=200$ and $V=-0.2$. Panels a), b) correspond to
the parameter sets in Fig.~2, respectively. Solid lines are analytically obtained; dashed
ones represent the numerical results.}
\end{figure}

On the other hand, we observe in Fig.~2 the widening of probability profile in comparison with one of \cite{NaBla}
at $V=0$ for the same environment parameters. Comparing, we also note the growing peaks of probability in Fig.~2a
which correspond to the zone with $2d_2=0.4$. It turns out that the attractors presence causes the pumping
effect, which depends on the local adsorption (determined by $r_n=1-2d_n$) and is already revealed in the
leading order ($\kappa=0$).

\section{Variance Computing in the Leading Order}

Variance $\langle x^2\rangle-\langle x\rangle^2$ is important RW characteristic, denoted here as
\begin{equation}
\Lambda(t)\equiv\int x^2\rmd\mu_t;\qquad
\langle x\rangle\equiv\int x\rmd\mu_t=0.
\end{equation}

In our approach, one has $\Lambda(t)=\Lambda_0(t)+O(\kappa)$,
\begin{equation}
\Lambda_0(t)=\int_{-\infty}^{\infty}[x(\xi)]^2\varphi(\xi,t)\rmd\xi.
\end{equation}
We focus on the properties of $\Lambda_0(t)$, corresponding to
the leading order approximation.

Although the function $x(\xi)$ can be presented similarly to (\ref{Yfunc}), it is convenient
to substitute it in terms of characteristic functions $\{\tilde\chi_n(\xi)\}$. We find that
\begin{eqnarray}
&&[x(\xi)]^2=\xi^2A_2(\xi)+|\xi|A_1(\xi)+A_0(\xi),\\
&&A_s(\xi)=\sum\limits_{n=1}^NA_{s,n}\tilde\chi_n(\xi);
\end{eqnarray}
where numeric coefficients are
\begin{eqnarray}
&&A_{0,n}=\left(c_n-\sqrt{d_n}b_{n-1}\right)^2,\\
&&A_{1,n}=2(\sqrt{d_n}c_n-d_nb_{n-1}),\\
&&A_{2,n}=d_n,\quad
c_n=\sum\limits_{m=1}^{n-1}\sqrt{d_m}(b_m-b_{m-1}).
\end{eqnarray}
Note that $A_2(\xi)$ coincides with diffusivity ${\tilde D}(\xi)$.

Combining the environment parameters $A_{s,n}$ and time-dependent integrals, we write
\begin{equation}\label{L0}
\Lambda_0(t)=2\sum\limits_{s=0}^2\sum\limits_{n=1}^NA_{s,n}[U_{s,V}(b_n,t)-U_{s,V}(b_{n-1},t)],
\end{equation}
where functions $U_{V,s}(b,t)$,
\begin{eqnarray}
&&\hspace*{-5mm}
U_{0,V}(b,t)=\frac{1}{4}\mathrm{erf}\left(\frac{b-Vt}{2\sqrt{t}}\right)
-\frac{\rme^{Vb}}{4}\mathrm{erfc}\left(\frac{b+Vt}{2\sqrt{t}}\right)\\
&&\hspace*{-5mm}
U_{1,V}(b,t)=-\frac{1}{2}\sqrt{\frac{t}{\pi}}\exp{\left(-\frac{(b-Vt)^2}{4t}\right)}
\nonumber\\
&&+\frac{1}{4V}\left[\mathrm{erf}\left(\frac{b-Vt}{2\sqrt{t}}\right)
+\rme^{Vb}\mathrm{erfc}\left(\frac{b+Vt}{2\sqrt{t}}\right)\right]
\nonumber\\
&&+\frac{1}{4}\left[tV\mathrm{erf}\left(\frac{b-Vt}{2\sqrt{t}}\right)
-b\rme^{Vb}\mathrm{erfc}\left(\frac{b+Vt}{2\sqrt{t}}\right)\right],\\
&&\hspace*{-5mm}
U_{2,V}(b,t)=\left(t+\frac{V^2t^2}{4}-\frac{1}{2V^2}\right)\mathrm{erf}\left(\frac{b-Vt}{2\sqrt{t}}\right)
\nonumber\\
&&-\frac{1}{2}\left(b+Vt+\frac{2}{V}\right)\sqrt{\frac{t}{\pi}}\exp{\left(-\frac{(b-Vt)^2}{4t}\right)}
\nonumber\\
&&-\frac{2-2Vb+V^2b^2}{4V^2}\rme^{Vb}\mathrm{erfc}\left(\frac{b+Vt}{2\sqrt{t}}\right),
\end{eqnarray}
determine the integrals for positive $b$ (or $|b|$):
\begin{equation}
\int^b_0\xi^s\varphi(\xi,t)\rmd\xi=U_{s,V}(b,t)-U_{s,V}(0,t).
\end{equation}

\begin{figure}[htbp]
\begin{picture}(80,45)
\put(-60,-114){\includegraphics[width=7.2cm,angle=0]{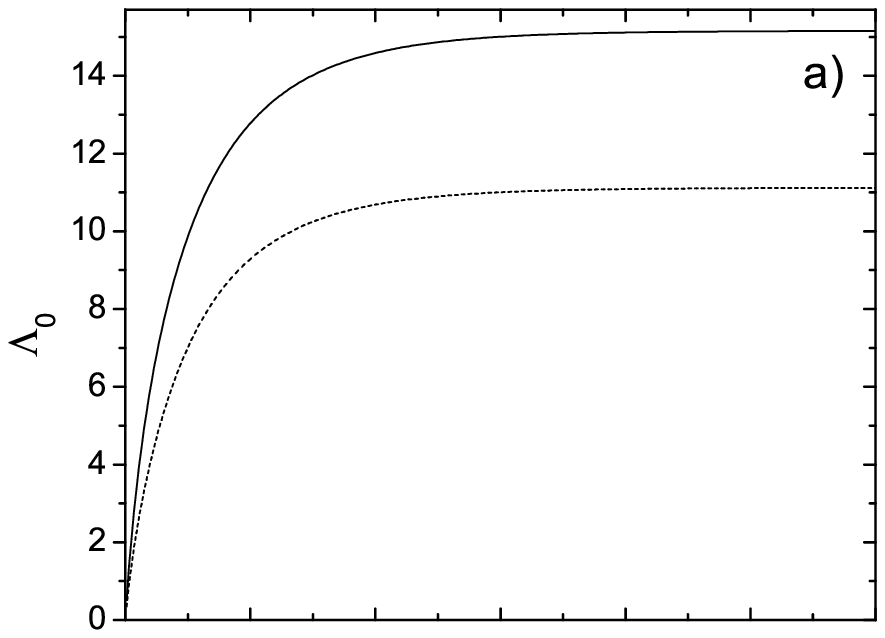}}
\put(-64.1,-259){\includegraphics[width=7.55cm,angle=0]{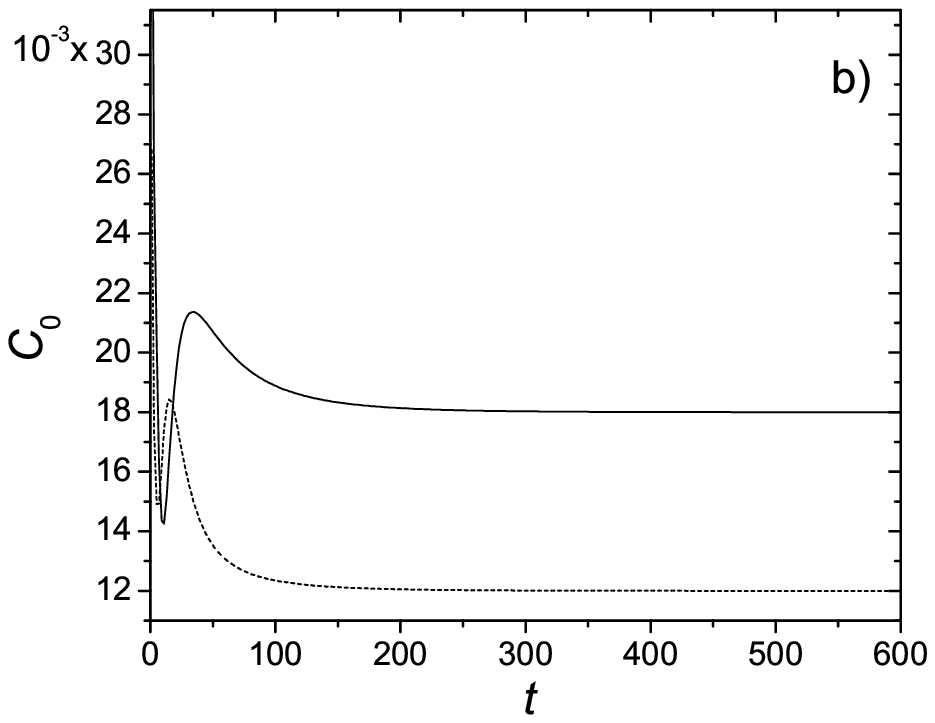}}
\end{picture}
\vspace*{90mm}
\caption{\label{fig4} 
Time dependencies of variance $\Lambda_0$ at $V=-0.2$ (a) and $C_0=\partial^2_t\Lambda_0/2$
at $V=0.2$ (b) for two sets of parameters. Solid curves correspond to $2d_1=0.6$,
$2d_2=0.4$, $2d_3=0.9$; dashed ones are for $2d_1=0.4$, $2d_2=0.9$, $2d_3=0.6$.}
\end{figure}

Questioning on behavior of $\Lambda_0(t)$ at large $t$, the answers turn out to be sufficiently
different for the models with $V<0$ and $V>0$, respectively. We can see that Fig.~4a shows an
existence of equilibrium and static limit $\langle x^2\rangle_\mathrm{eq}=\lim_{t\to\infty}\Lambda_0(t)$
at $V<0$, while Fig.~4b indicates the ballistic regime for $V>0$ with $\langle x^2\rangle\sim C_0t^2$ at
$t\to\infty$. Moreover, the environment structure, given by the different sets of parameters $\{d_n\}$,
does not affect these tendencies.

Assuming an attractor presence at $x=0$ ($V<0$) and taking $t\to\infty$ in (\ref{L0}),
one has
\begin{eqnarray}
\langle x^2\rangle_\mathrm{eq}&=&\frac{2d_1}{V^2}
+\frac{1}{V^2}\sum\limits_{n=1}^{N-1}\rme^{-|V|b_n}\left[V^2(A_{0,n+1}-A_{0,n})\right.\nonumber\\
&&+|V|(A_{1,n+1}-A_{1,n})(1+|V|b_n)\nonumber\\
&&\left.+(d_{n+1}-d_n)(2+2|V|b_n+V^2b^2_n)\right].\label{Leq}
\end{eqnarray}

It follows that the maximum values of $\Lambda_0(t)$ for models in Fig.~4a are
$\langle x^2\rangle_{\mathrm{eq},1}\approx15.16$ and $\langle x^2\rangle_{\mathrm{eq},2}\approx11.12$,
when $(2d_1/V^2)_1=15$ and $(2d_1/V^2)_2=10$, respectively. Thus, points $x=\pm\sqrt{\langle x^2\rangle_\mathrm{eq}}$
lie inside the first zone $(-a_1;a_1)$ for both models.

Focusing on the case of $V>0$, a time asymptotic of $\Lambda_0(t)$ is determined by the term containing
$U_{2,V}(b_N,t)-U_{2,V}(b_{N-1},t)$ at $b_N\to\infty$. Thus, $\langle x^2\rangle=d_NV^2t^2$ at $t\to\infty$.
This formula is valid, when the PDF peaks are far from region $x\in(-a_{N-1};a_{N-1})$ of the basic environment
structure presence (see Fig.~1). Thus, we arrive at the asymptotic values of $C_0$ in Fig.~4b: $(d_3V^2)_1=0.018$
and $(d_3V^2)_2=0.012$.

We would like to note that the velocity correlator ${\tilde C}=C_0-d_NV^2$ reproduces here the typical properties
of $C$ at $V=0$ from \cite{NaBla} for the same environment parameters $\{a_n;d_n\}$. We observe similarity in
a sign alternating ${\tilde C}$ at small $t$ for the model with $2d_3=0.9$, while the regime ${\tilde C}>0$
is preserved during the whole evolution for the model with $2d_3=0.6$.

Note that the ${\tilde C}$'s sign allows us to classify a system behavior. Stages with ${\tilde C}>0$ correspond,
as usual, to superdiffusion. On the other hand, the regime with ${\tilde C}<0$ reveals the transient subdiffusion,
caused by particle capture for a short time by zones with relatively small~$d_n$.

Thus, the adsorption effect of each zone, determined by $r_n=1-2d_n$, lasts important at $V>0$, although
a magnitude and time intervals of ${\tilde C}$'s variations depend on $V$. Moreover, the models with $V>0$
give rise a possibility to investigate the influence of the distant zones in comparison with the models at
$V=0$ during the same time $t$. On contrary, the case of $V<0$ does not allow here to overstep actually
the first zone.

\section{Discussion}

We consider an asymmetric RW model in a one-dimensional space, densely covered by the finite-sized zones
$x\in(-a_n,-a_{n-1})\cup(a_{n-1},a_n)$, $n=\overline{1,N}$, with the properties specified, what allows us
to investigate an affect of inhomogeneities in chaotic systems of a different nature.
Although a number of the RW characteristics can be extracted from numerical simulations, a deeper analysis
requires an analytical description which is based in continuous limit on the advection-diffusion equation
at the large total number of steps $t$, associated with a time.

Our model, initially formulated in the terms of probabilities, is designed also to include the familiar
problems of RW with various barriers \cite{Ascher60,Lehner,Gupta,Tanner,Percus,BHMST,NFJH} into the concept
of the heterogeneous environment. At this time, these stimulate us by their predictions, methods and the
unsolvable problems there.

Diffusion and advection are main and competing processes which we account for. Their parameters (diffusion
coefficients $0<d_n\leq1/2$ and drift velocities $v_n=V\sqrt{d_n}$), varying from zone to zone, permit
to reveal also an adsorption in the bulk (determined by $r_n=1-2d_n$) and an action of long range external
field $V$. Sometimes, the latter notions are convenient for physical treatment of the observed phenomena.

Widening the model understanding, we can also relate the diffusion coefficient to an effective
walker mass $m_n=1/2d_n$ for each zone. Then, thinking about macroscopic particle ensemble,
the mass variations might be interpreted as the result of geometrically dependent interaction among
particles which is not specified here but leads to the emergence of different states, confined
within the zones.

Before summarizing the results, we note briefly a role of advection, which contributes the directed
motion into RW and is induced by a global asymmetry parameter $-1<V<1$, determining the preference
between the left and right directions.

Precisely, the advection is involved here by means of the space-dependent velocity $\varepsilon_xV\sqrt{D(x)}$,
where $\varepsilon_x\equiv\mathrm{sign}(x)$ makes the RW root point $x=0$ to play an attractive or repulsive
role, depended on the $V$'s sign. The case of $V<0$ means that the point $x=0$ attracts a walker.
On contrary, at $V>0$, we have a repulsor at $x=0$, sending a walker from there in one of mutually opposite
directions of axis~$x$. 

Technically, such a definition preserves the model symmetry under coordinate inversion and leads to
$\langle x\rangle=0$ for any $V$.

Thus, the diffusion and advection processes under our assumptions on the environment piece-wise structure determine
a probability to find a walker at space-time point. Finding a probability distribution function (PDF) from
the advection-diffusion differential equation, the parameter constancy {\it almost} everywhere looks as a
crucial condition of the problem solvability. Indeed, the geometrical data are easily concentrated in new
spatial variable $\xi(x)$ what leads to equation for homogeneous environment with the singular terms,
corresponding to the residual interface effects and containing the Dirac $\delta$-function and its derivatives.
Although this contradicts to a probabilistic meaning of the quantities involved, it is a sequel of
using the distributions in continuous limit. The lattice RW simulations have no singularities. However,
a receipt of obtaining a PDF by accounting for the interface contribution has to be found yet. It may be
resolved exactly by constructing the solutions with a gap \cite{Powels} which are not considered here.

To obtain a PDF, we use a formal series in a switching parameter $\kappa\sim1$, controlling the surface
effects in equation. The leading term by $\kappa^0$ gives us a basic and normalized solution for the bulk,
which is equal in the terms of $\xi$ to the fundamental solution of homogeneous system. Nevertheless,
the resulting PDF of $(x,t)$ reflects an environment complexity and the drift presence. We have also found
a solution with the surface correction by $\kappa^1$, which is applicable at $V\geq0$ and
$|d_{n+1}-d_n|\ll1$ as shown. We should use additional restrictions to limit the correction magnitude,
which is not suppressed by $\kappa$.

Although such an approach is already exploited and justified by numerical simulations in \cite{NaBla}
without drift, the advection inclusion changes considerably the system dynamics and the PDF form,
resulting in new outcomes.

To describe analytically the RW, we also compute the probability function in the linear approximation in $\kappa$
and the variance dependence on $t$ in the leading order, neglecting the interface term.

For any $N$, we reveal the variance time asymptotic:
\begin{equation}
\langle x^2\rangle\sim t^{1+\mathrm{sign}(V)},\qquad t\to\infty.
\end{equation}
This tendency is independent on the environment structure with $d_n>0$ and also happens for
uniform models as is already known.

However, an environment complexity leads at finite $t$ to effective power law $\langle x^2\rangle\sim t^\alpha$
with intermediate values of exponent $\alpha$, indicating a transient anomalous diffusion. It is clearly seen at
$V\geq0$.

Showing it, we appeal to the models with a three-zone environment, given by the parameter sets
from~\cite{NaBla}. Furthermore, it is convenient to study the diffusion regimes by means of
the velocity autocorrelation function
\begin{equation}
\tilde C(t)=\frac{1}{2}\frac{\rmd^2\langle x^2\rangle}{\rmd t^2}-v_N^2,\qquad
v_N=V\sqrt{d_N}.
\end{equation}
Then, super/sub-diffusion processes correspond to ${\rm sign}({\tilde C}(t))=1/(-1)$, respectively.
Using that, such time intervals are found in Fig.~4b.

It is important to note that ${\tilde C}$ for fixed set $\{a_n;d_n\}$ indicates a similar (anomalous) behavior
at $V=0.2$ and $V=0$ (see \cite{NaBla}). Although a possibility to observe an affect of distant zones at $V>0$
is higher than at $V=0$ for the same $t$.

Comparing the probability profiles at $V=0.2$ and $V=0$ (see \cite{NaBla}), we can see that the adsorption
property of zones is intensified in the advection presence (see Fig.~2). Such a pumping effect
looks surprisingly because the advection and the diffusion are competing.

At negative values of $V$ one has the PDF relaxation to a steady state distribution when $t\to\infty$.
Although a walker is attracted to the origin $x=0$, diffusion prevents a collapse with $\langle x^2\rangle=0$.
However, there is no manifestation of all zones.

Note finally that a perspective is analytical description of the model with
independent local velocities $\{v_n\}$, extending the parameter set up to $\{a_n;d_n;v_n\}$.
Formalism, accounting for local fluctuations $\delta{\cal D}(x)\equiv{\cal D}(x)-D(x)$
of smooth-varied diffusivity ${\cal D}(x)$, where $D(x)=\sum_{\{n\}} d_n\chi_n(x)$,
could be developed.

\vskip3mm

Authors are indebted for partial support of this work by Department of Physics and Astronomy of
NAS of Ukraine.




\begin{thebibliography}{99}

\bibitem{NaBla}
A.V.~Nazarenko and V.~Blavatska, J. Phys. A: Math. Theor. {\bf 50}, 185002 (2017).

\bibitem{Ascher60}
M.~Ascher, Math. Comp. {\bf 14}, 346 (1960).

\bibitem{Lehner}
G.~Lehner, Ann. Math. Statist. {\bf 34}, 405 (1963).

\bibitem{Gupta}
H.S.~Gupta, J. Math. Sci. {\bf 1}, 18 (1966).

\bibitem{Tanner}
J.E.~Tanner, J. Chem. Phys. {\bf 69}, 1748 (1978).

\bibitem{Percus}
O.E.~Percus and J.K.~Percus, SIAM J. Appl. Math. {\bf 40}, 485 (1981);
O.E.~Percus, Adv. Appl. Prob. {\bf 17}, 594 (1985).

\bibitem{BHMST}
P.S.~Burada, P.~H\"anggi, F.~Marchesoni, G.~Schmid, P.~Talkner, Chem. Phys. Chem. {\bf 10}, 45 (2009).

\bibitem{NFJH}
D.S.~Novikov, E.~Fieremans, J.H.~Jensen, J.A.~Helpern, Nat. Phys. {\bf 7}, 508 (2011).

\bibitem{Powels}
J.G.~Powels, M.J.D.~Mallett, G.~Rickayzen, W.A.B.~Evans, Proc. R. Soc. Lond. A {\bf 436}, 391 (1992).

\bibitem{RWbook}
See  e.g. M.F.~Shlesinger and B.~West (ed)  {\em Random Walks and their Applications in the Physical and Biological Sciences}
(AIP Conf  Proc  vol  109) (AIP, New York, 1984);
F.~Spitzer  {\em Principles of Random Walk} (Springer, Berlin, 1976).

\bibitem{Havlin87}
S.~Havlin and D.Ben~Abraham, Phys. Adv. {\bf 36}, 695 (1987).

\bibitem{MGCh}
R.~Metzler, J.-H.~Jeon, A.G.~Cherstvy, Acta BBA-Biomembr. {\bf 1858}, 2451 (2016).

\bibitem{Pat}
S.V.~Patankar {\em Numerical Heat Transfer and Fluid Flow} (McGraw-Hill, New York, 1980).

\bibitem{GPSG}
J.S.~Perez~Guerro, L.C.G.~Pimentel, T.H.~Skaggs, M.Th. van Genuchten, Int. J. Heat Mass Trans. {\bf 52},
3297 (2009).

\bibitem{KJK}
A.~Kumar, D.~Kumar~Jaiswal, N.~Kumar, J. Hydrol. {\bf 380}, 330 (2010);
A.~Kumar, D.~Kumar~Jaiswal, R.R~Yadav, IOSR J. Math. {\bf 2}, 1 (2012).

\bibitem{Singh}
R.N.~Singh, J. Ind. Geophys. Union {\bf 17}, 117 (2013).

\end{thebibliography}
\end{document}